\documentclass[10pt,letterpaper]{article}
\usepackage{opex3}

\begin{document}
\title{Deterministic design of wavelength scale, ultra-high Q photonic crystal nanobeam cavities}
\author{Qimin Quan$^*$ and Marko Loncar}
\address{School of Engineering and Applied Sciences, Harvard University, Cambridge, MA 02138, USA}
\email{$^*$quan@fas.harvard.edu}

\begin{abstract}
Photonic crystal nanobeam cavities are versatile platforms of interest for optical communications, optomechanics, optofluidics, cavity QED, etc. In a previous work \cite{quan10}, we proposed a deterministic method to achieve ultrahigh \emph{Q} cavities. This follow-up work provides systematic analysis and verifications of the deterministic design recipe and further extends the discussion to air-mode cavities. We demonstrate designs of dielectric-mode and air-mode cavities with $Q>10^9$, as well as cavities with both high-\emph{Q} ($>10^7$) and high on-resonance transmissions ($T>95\%$).
\end{abstract}


\bibliography{amsplain}

\begin{thebibliography}{99}
\bibitem{note1} Quality factor is defined as $Q=\omega_0 \frac{\mathrm{Energy\, stored}}{\mathrm{Power\, loss}}$, and mode volume is defined as $V=\int
\mathrm{d}V\epsilon|\mathbf{E}|^2/[\mathbf{\epsilon}|\mathbf{E}|^2]_{\mathrm{max}}$.
\end{thebibliography}

\section{Introduction}
High quality factor (\emph{Q}), small mode volume (\emph{V})\cite{note1} optical cavities provide powerful means
for modifying the interactions between light and matter\cite{vahala03re}, and have
many exciting applications including quantum information
processing\cite{obrien09}, nonlinear optics\cite{leuthold10}, optomechanics\cite{thourhout10}, optical trapping\cite{greier03} and optofluidics\cite{psaltis06}. Photonic crystal
cavities (PhC)\cite{Yablonovitch87}\cite{John87} have demonstrated
numerous advantages over other cavity geometries due to their wavelength-scale mode volumes
and over-million \emph{Q}-factors\cite{Foresi97}-\cite{Deotare08}. Although small mode volumes of PhC cavities can be easily achieved by design, ultrahigh \emph{Q} factors are typically obtained using extensive parameter search and optimization. In a previous work\cite{quan10}, we proposed a deterministic method to design an ultrahigh \emph{Q} PhC nanobeam cavity and verified our designs experimentally. The proposed method does not rely on any trial-and-error based parameter search and does not require any hole shifting, re-sizing and overall cavity re-scaling. The key design rules we proposed that result in ultrahigh \emph{Q} cavities are (i) zero cavity
length ($L=0$), (ii) constant length of each mirror ('period'=$a$) and (iii) a Gaussian-type of field attenuation profile, provided by linear increase in the mirror strength.

In this follow-up work, we provided numerical proof of the proposed principles, and systematically optimized the design recipe to realize a radiation limited cavity and waveguide coupled cavity respectively. Furthermore, we extended the recipe to the design of air-mode cavities, whose optical energies are concentrated in the low-index region of the structure.

Nanobeam cavities have recently emerged as a powerful alternative to the slab-based 2-D PhC cavities\cite{Noda03}-\cite{noda08}. Nanobeams can achieve \emph{Q}s
on par with those found in slab-based geometries, but in much smaller footprints, and are the most natural geometries for integration with waveguides\cite{Notomi08}-\cite{Deotare08}. Our deterministically designed cavities have similar structures to the mode-gap cavity proposed by Notomi et al. \cite{Notomi08}, and later demonstrated experimentally by Kuramochi et al.\cite{kuramochi10}, as well as our own work\cite{quan11}. We note that the same design principle discussed here could be directly applied to realize ultra-high \emph{Q} cavities based on dielectric stacks that are of interest for realization of vertical-cavity surface emitting lasers (VCSELs) and sharp filters. Finally, it is important to emphasize that while our method is based on the framework of Fourier space analysis\cite{Vuckovic02,Katik02,Dirk05}, alternative approach, based on phase-matching between different mirror segments, could also be used to guide our design, as well as to explain the origin of deterministic ultra-high \emph{Q}-factors in our devices\cite{lalanne01,lalanne04}.

\section{Numerical Verification of the Deterministic Design Approach}
\begin{figure}
\centering
\includegraphics[width=12cm,height=2.5cm]{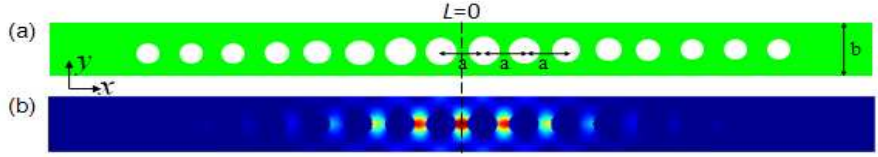}
\caption{\small (a) Schematic of the proposed nanobeam cavity. (b) FDTD simulation of the energy density distribution in the middle plane of the nanobeam cavity. }\label{schematics}
\end{figure}

\begin{figure}
\centering
\includegraphics[width=11cm,height=16cm]{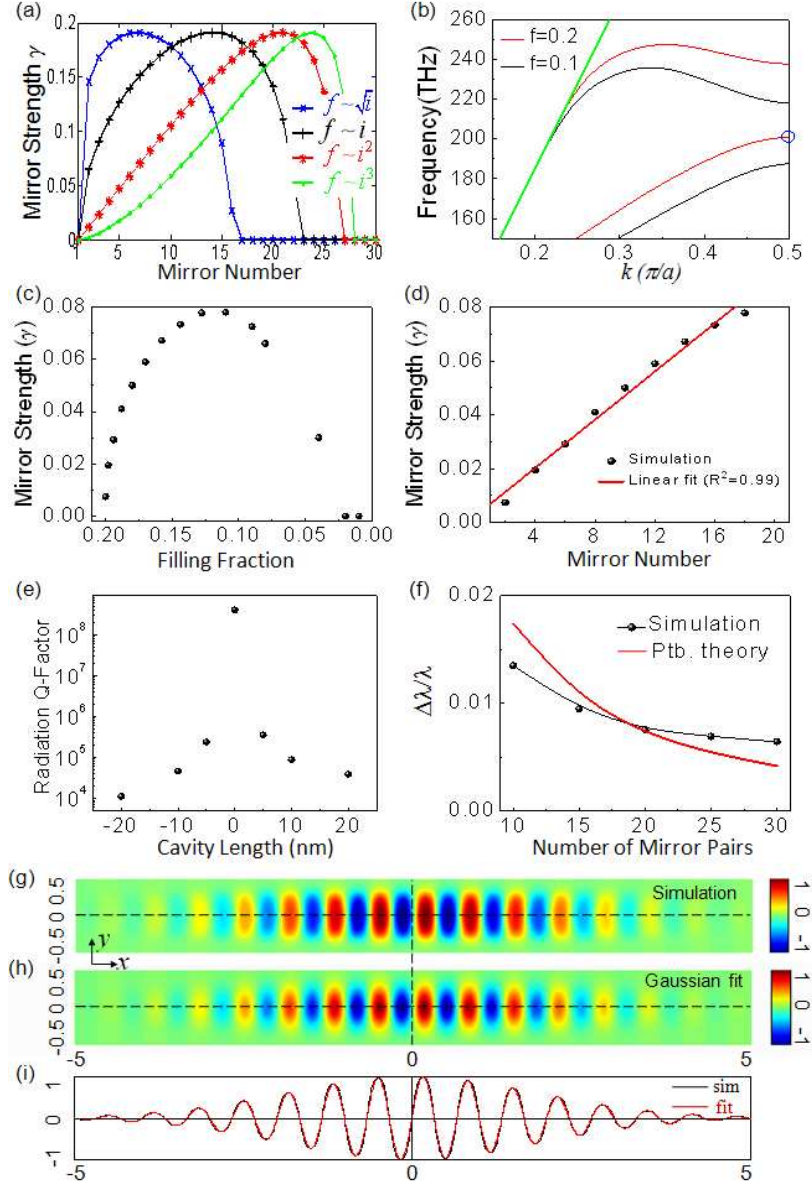}
\caption{\small (a) Mirror strengths of each mirror segment under different tapering profiles obtained from the plane wave expansion method ('1' indicates the mirror segment in the center of the cavity). (b) Band diagram of the TE-like mode for $f=0.2$ and $f=0.1$. The green line indicates the light line. The circle indicates the target cavity resonant frequency. (c) Mirror strengths at different filling fractions from the 3D band diagram simulation. (d) Mirror strengths as a function of mirror number after quadratic tapering. (e) Radiation-\emph{Q} factors when nanobeam cavities have different cavity lengths between the two Gaussian mirrors. (f) Resonances of the cavities that have different total number of mirror pair segments in the Gaussian mirror, and their
deviations from the dielectric band-edge of the central mirror segment, obtained from FDTD simulation and perturbation theory. (g) $H_z$ field distribution on the surface right above the cavity, obtained from 3D FDTD simulation. The structure has dimension of $a=0.33\mu m, b=0.7\mu m$, the first 20 mirror segments (counted from the center) have $f$ varies from 0.2 to 0.1, followed by 10 additional mirror segments with $f=0.1$. (h) $H_z$ field distribution on the surface right above the cavity, obtained from the analytical formula $H_z= \sin(\frac{\pi}{a} x)\exp(-\sigma x^2)\exp(-\xi y^2)$, with $a=0.33\mu m,
\sigma=0.14, \xi=14$. (i) $H_z$ field distribution along the dashed line in (g)\&(h). Length unit in (g)-(i) is $\mu$m.}\label{validation1}
\end{figure}

In this section, we use finite-difference time-domain (FDTD) simulations to systematically study the design principles proposed in \cite{quan10}. We consider a TE-polarized dielectric-mode cavity, i.e. $E_y, H_x, H_z$ are the major field components, and the energy density is concentrated in high index material (silicon in our case). Fig.~\ref{schematics}(a) shows the schematics of the proposed cavity structure\cite{quan10}. It consists of an array of air-holes in decreasing radii, etched into a ridge waveguide. The hole-to-hole distances are constant ("periodicity"). The structure is symmetric with respect to the dashed line in Fig.~\ref{schematics}(a). In contrast to the majority of other cavity designs, current structure has no additional cavity length inserted between the two mirrors (\emph{L}=0), that is the distance between the two central holes is the same as that of the rest of the structure ($a$). This design minimizes the cavity loss and the mode volume simultaneously. The cavity loss is composed of the radiation loss into the free space (characterized by $Q_{\mathrm{rad}}$) and the coupling loss to the feeding waveguide ($Q_{\mathrm{wg}}$). $Q_{\mathrm{wg}}$ can be increased simply by adding more gratings along the waveguide. $Q_{\mathrm{rad}}$ can be increased by minimizing the spatial Fourier harmonics of the cavity mode inside the lightcone. The optical energy is concentrated in the dielectric region in the middle of the cavity (Fig.~\ref{schematics}(b)). In order to achieve a Gaussian-type attenuation, we proposed to use a linearly increasing mirror strength along the waveguide\cite{quan10}, which was achieved by tapering the hole radii. To analyze the ideal tapering profile, in this work, we first use plane wave expansion method and then verify the results with 3D FDTD simulations. The dielectric profile of the structure in the middle plane of the cavity can be expressed as
\begin{equation}
\frac{1}{\epsilon(\mathbf{\rho})}=\frac{1}{\epsilon_{\mathrm{Si}}}+(1-\frac{1}{\epsilon_{\mathrm{Si}}})S(\mathbf{\rho})
\end{equation}
with
$$
S(\rho) = \left\{
\begin{array}{lr}
1& |\mathbf{\rho-r}_j|\le R\\
0& |\mathbf{\rho-r}_j|> R
\end{array}
\right.
$$
$\mathbf{r}_j=j\cdot a \hat{x}$, $a$ is the period, and $j=\pm1,\pm2...$ are integers. $R$ is the radius of the hole. Using plain wave expansion method\cite{sakoda} in the beam direction ($\hat{x}$),
\begin{equation}
\frac{1}{\epsilon(x)}=\kappa_0+\kappa_1 e^{i{G}{x}}+\kappa_{-1} e^{-i{G}{x}}+...
\end{equation} where $G=2 \pi/a$. The zeroth ($\kappa_0$) and first ($\kappa_1$) order Fourier components can be expressed as \cite{sakoda}
\begin{eqnarray}
\kappa_0=f+\frac{1-f}{\epsilon_{\mathrm{Si}}}\\
\kappa_1=2f(1-\frac{1}{\epsilon_{\mathrm{Si}}})\frac{ J_1( G R)}{G R}
\end{eqnarray}
$J_1$ is the first order Bessel function. Filling fraction $f=\pi R^2/ab$ is the ratio of the area of the air-hole to the area of the unit cell.

The dispersion relation can be obtained by solving the master equation \cite{joannopoulosbook}:
\begin{equation}
\frac{c^2}{\epsilon(x)}\frac{\partial^2 E}{\partial x^2}=\frac{\partial^2 E}{\partial t^2}
\end{equation}
Inside the bandgap, the wavevector ($k$) for a given frequency ($\omega$) is a complex number, whose imaginary part denotes the mirror strength ($\gamma$). For solutions near the band-edge, of interest for high-\emph{Q} cavity design\cite{quan10}, the wavevector $k=(1+i\gamma)\pi /a$, and frequency $\omega=(1-\delta)
\sqrt{\kappa_0}\pi c/a$. Substituting this into the master equation,
we obtained
$\delta^2+\gamma^2=\kappa_1^2/4\kappa_0^2$. The cavity resonance
asymptotes to the dielectric band-edge of the center mirror segment:
$w_{\mathrm{res}}\rightarrow(1-\kappa_1^{j=1}/2\kappa_0^{j=1})
\sqrt{\kappa_0^{j=1}}\pi c/a$ ($j$ represents the
$j^\mathrm{th}$ mirror segment counted from the center), at which point the mirror strength $\gamma^{j=1}=0$. $\gamma$ increases with $j$. Fig.~\ref{validation1}(a) shows $\gamma-j$ plot for different tapering profiles, which is the dependence of the filling fraction ($f$) on the index of mirror segment ($j$) starting from the center ($j=1$). It can be seen that quadratically tapering profile results in linearly increasing mirror strengths, needed for Gaussian field attenuation\cite{quan10}. To verify this, we numerically solved the band diagram (Fig.~\ref{validation1}(b)) and obtained $\gamma-f$ relation in Fig.~\ref{validation1}(c). As shown in Fig.~\ref{validation1}(d), linearly increasing mirror strength is indeed achieved after quadratic tapering.

Next, with the optimized tapering profile, the cavity is formed by putting two such mirrors back to back, leaving a cavity length \emph{L} in between (Fig.~\ref{schematics}(a)). Fig.~\ref{validation1}(e) shows the simulated \emph{Q}-factors for various \emph{L}s.  Highest $Q_{\mathrm{rad}}$ is achieved at zero cavity length (\emph{L}=0), which supports the prediction in \cite{quan10} based on 1D model.

Third, we verify that the cavity mode has a Gaussian-type attenuation profile. Fig.~\ref{validation1}(g) shows the $H_{z}$-field distribution in the plane right above the cavity, obtained from 3D FDTD simulation. As shown in Fig.~\ref{validation1}(h), this field distribution can be ideally fitted with $H_z=\sin(\pi x/a)\exp(-\sigma x^2)\exp(-\xi y^2)$, with $a=0.33$, $\sigma=0.14$ and $\xi=14$. The fitted value $a$ agrees with the "period", and $\sigma$ agrees with that extracted value from Fig.~\ref{validation1}(d): $\sigma=\frac{\mathrm{d}\gamma}{\mathrm{d} x}\frac{\pi}{a} =0.13$. Fig.~\ref{validation1}(i) shows $H_z$ distribution along the dashed line in Fig.~\ref{validation1}(g)\&(h). Therefore, we conclude that zero cavity length, fixed periodicity and a quadratic tapering of the filling fraction results in a Gaussian field profile, which leads to a high-\emph{Q} cavity\cite{quan10}.

Finally, as we have pointed out in \cite{quan10}, current method results in a cavity whose resonance is asymptotically approaching the dielectric band-edge frequency of the central mirror segment (circled in Fig.~\ref{validation1}(b)). The deviation from the band-edge frequency can be calculated using perturbation theory\cite{joannopoulosbook}\cite{johnson02}:
\begin{equation}
\frac{\delta \lambda}{\lambda}=\frac{\int \delta \epsilon |\textbf{E}_\parallel|^2 - \delta (\epsilon^{-1}) |\textbf{D}_\perp|^2
dV}{2\int \epsilon |\textbf{E}|^2 dV} \label{perturbation}
\end{equation}
$\textbf{E}_\parallel$ is the component of $\textbf{E}$ that is parallel to the side wall surfaces of the holes and $\textbf{D}_\perp$ is the component of $\textbf{D}$ that is perpendicular to the side wall surfaces of the holes. Under Gaussian distribution, the major field component $D_y=\cos (\pi/ax) \exp(-\sigma x^2)\exp(-\xi y^2)$,
$\delta \epsilon$ perturbation occurs at $x = \pm(j+1/2) a \pm
R_j$, where $R_j=\sqrt{f_j a b/\pi}$ denotes the radius of the $j^{\mathrm{th}}$ hole (counted the center), with $j$=1,2...$N$, $N$ is the total number of mirror segments at each side. Since the cavity mode has a Gaussian profile, $1/\sqrt{\sigma}$ characterizes the effective length of the cavity mode, and scales linearly with $N$, with a nonzero intercept due to diffraction limit. For large $N$, the intercept can be neglected, and thus $\sigma_N=\sqrt{20\times0.14^2/N}$. Plug the perturbation induced by the quadratic tapering from $f=0.2$ to $f=0.1$ into Eqn.~\ref{perturbation}, the frequency offset $\delta\lambda/\lambda$ v.s $N$ can be obtained. Fig.~\ref{validation1}(f) shows the frequency offset for different total number of mirror pairs ($N$), calculated from the perturbation theory, as well as using FDTD simulations. It can be seen that the deviation decreases
as the number of modulated mirror segments increases, and is
below 1\% for $N>15$.

\begin{figure}
\centering
\includegraphics[width=12cm,height=7cm]{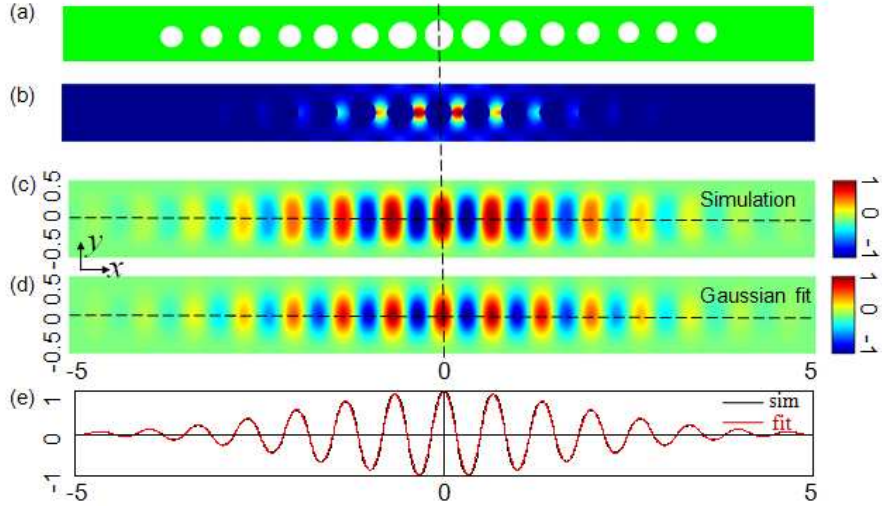}
\caption{\small (a) Schematic of the Gaussian nanobeam cavity, with an air hole in the symmetry plane (dashed line). (b) Energy distribution in the middle plane of the cavity obtained from 3D FDTD simulation. (c)\&(d) $H_z$ field distribution on the surface right above the cavity: (c) is obtained from 3D FDTD simulation and (d) is obtained from the analytical formula $H_z= \cos(\frac{\pi}{a} x)\exp(-\sigma x^2)\exp(-\xi y^2)$, with $a=0.33\mu m,
\sigma=0.14, \xi=14$. (e) $H_z$ field distribution along the dashed line in (c)\&(d). Length unit: $\mu$m.}\label{validation2}
\end{figure}

Therefore, we verified that an ultrahigh-\emph{Q}, dielectric-mode cavity resonant at a target frequency can be designed using the following algorithm:

(i) Determine a target frequency. For example in our case we want $f_{\mathrm{target}}= 200$THz. Since the  cavity  resonance is  typically 1\% smaller than the dielectric band-edge of the central segment, estimated using the perturbation theory, we shift-up  the target frequency by 1\%, i.e. $f_{\mathrm{adjusted}}= 202$THz.

(ii) Pick the thickness of the nanobeam: this is often pre-determined by the choice of the wafer. For example,  in our case, the thickness of the nanobeam is 220nm, determined by the thickness of the device layer  of our silicon-on-insulator (SOI) wafer.

(iii) Choose periodicity according to  $a=\lambda_0/2n_{\mathrm{eff}}$, where $n_{\mathrm{eff}}$ is effective mode index of the cavity and can be estimated by numerical modeling of a strip waveguide that nanobeam cavity is based on. However, we found that the absolute value of the periodicity is not crucial in our design, as long as there exists a bandgap. Therefore, we pick $n_{\mathrm{eff}}=2.23$, which is a median value of possible effective indices in the case of free standing silicon nanobeam ($n_{\mathrm{eff}} \in [1, 3.46]$). This results in $a=330$nm.

(iv) Set nanobeam width. Large  width increases the effective index of the cavity mode, pulls the mode away from the light line, thus reducing the in plane radiation loss. On the other hand, a large beam width will allow for higher order modes with the same symmetry as the fundamental mode of interest. Using band diagram simulations, we found that the width of 700nm is good trade-off between these two conditions (Fig.~\ref{validation1}(b)).

(v) Set the filling fraction of the first mirror section such that its dielectric band-edge is at  the adjusted frequency: 202THz in our case. Band diagram calculations based on single unit cells are sufficient for this analysis. We found that  an optimal filling fraction in our case is $f_{\mathrm{start}}=0.2$ (Fig.~\ref{validation1}(b)).

(vi) Find the filling fraction that produces the maximum mirror strength for the target frequency. This involves calculating the mirror strength for several filling fractions (Fig.~\ref{validation1}(c)), each of which takes one or two minutes on a laptop computer. In our case we found that $f_{\mathrm{end}}=0.1$.

(vii) Pick the number of mirror segments ($N_{\mathrm{}}$) to construct the Gaussian mirror: we found that $N_{\mathrm{}}\geq15$ (on each side) are generally good to achieve high radiation-\emph{Q}s.

(viii) Create the Gaussian mirror by tapering the filling fractions quadratically from $f_{\mathrm{start}}$ (=0.2 in our case) to $f_{\mathrm{end}}$ (=0.1) over the period of $N_{\mathrm{}}$ segments. From the above analysis, the mirror strengths can be linearized through quadric tapering (Fig.~\ref{validation1}(d)).

(ix) Finally, the cavity is formed by putting two Gaussian mirrors back to back, with no additional cavity length in between ($L=0$). To achieve a radiation-limited cavity ($Q_{\mathrm{wg}}>>Q_{\mathrm{rad}}$), 10 additional mirrors with the maximum mirror strength are placed on both ends of the Gaussian mirror. We will show in the next section, no additional mirrors are needed to achieve a waveguide-coupled cavity ($Q_{\mathrm{rad}}>>Q_{\mathrm{wg}}$).

Besides the structure that were proposed in \cite{quan10} (Fig.~\ref{schematics}), the alternative structure which has the air-hole in the symmetry plane, as shown in Fig.~\ref{validation2}(a), also satisfies (i)-(ix). Both structures result in dielectric-mode cavities, since the bandgap of each mirror segments red-shifts away from the center of the cavity, and thus a potential well is created for the dielectric band-edge mode of the central segment. The difference is that the energy maximum in the air-hole centered cavity is no longer located in the middle of the structure, but instead in the dielectric region next to the central hole (Fig.~\ref{validation2}(b)). Fig.~\ref{validation2}(c) shows the $H_z$ field profile in the plane right above the cavity, obtained from FDTD simulation. Fig.~\ref{validation2}(d) shows the fitted field profile using the exactly the same parameters that were used in the original structure shown in Fig.~\ref{validation1}(g)-(i), but with sine function replaced by cosine function. Fig.~\ref{validation2}(e) shows the $H_z$ distribution along the dashed line in Fig.~\ref{validation2}(c)\&(d).

\begin{figure}
\centering
\includegraphics[width=12cm,height=12cm]{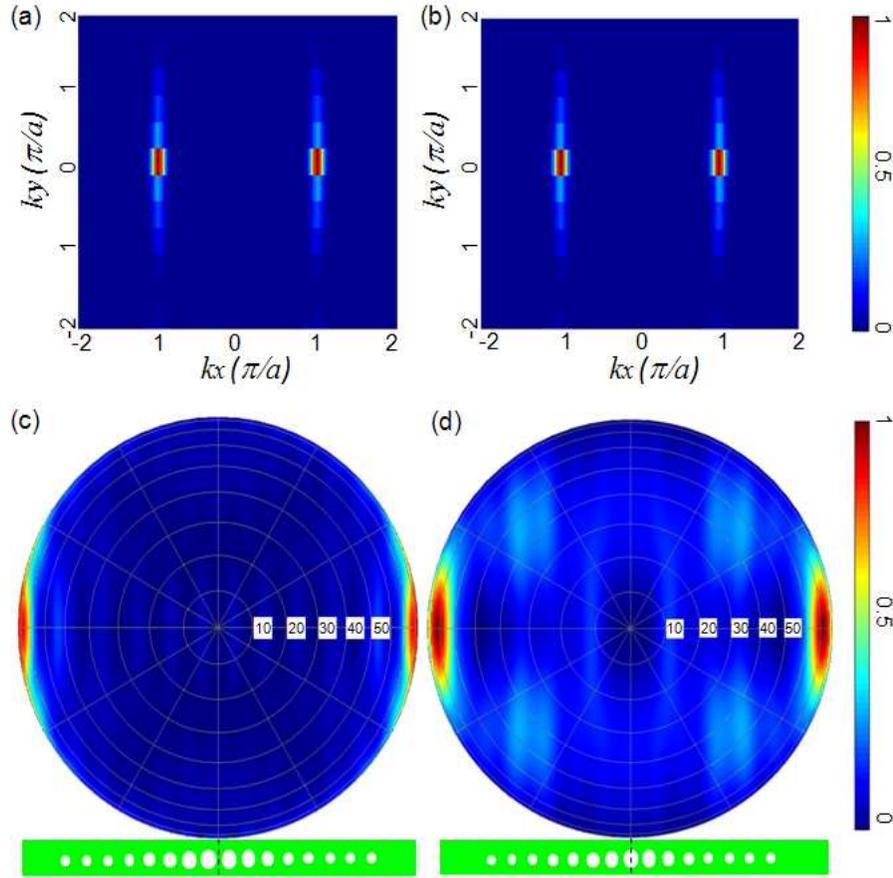}
\caption{\small (a)\&(b) The distribution of the spatial Fourier components of the cavity mode, obtained from 3D FDTD simulation: (a) for the $H^{odd}_z$ cavity and (b) for the $H^{even}_z$ cavity respectively. (c)\&(d) The far field profile of the cavity mode obtained from 3D FDTD simulation: (c) for the $H^{odd}_z$ cavity and (d) for the $H^{even}_z$ cavity respectively. The inset cavity structure shows the orientation of the waveguide direction in (c)\&(d). Dashed line indicates the symmetry plane. }\label{validation3}
\end{figure}

Armed with the analytical field profile of the cavities: $H^{odd}_z(x)=\sin(\pi x/a)\exp(-\sigma x^2)$ and $H^{even}_z(x)=\cos(\pi x/a)\exp(-\sigma x^2)$, we can obtain the radiation losses and and far fields of the cavities using the Fourier space analysis\cite{Vuckovic02}. The Fourier transforms can be analytically obtained $\mathrm{FT}(H^{odd}_z)=(\exp(-(k+\pi/a)^2/4\sigma)-\exp(-(k-\pi/a)^2/4\sigma))/i\sqrt{8\sigma}$ and $\mathrm{FT}(H^{\mathrm{even}}_z)=(\exp(-(k+\pi/a)^2/4\sigma)-\exp(-(k-\pi/a)^2/4\sigma))/\sqrt{8\sigma}$. Under $\sigma a^2<<1$, both distributions have their Fourier components strongly localized at $k=\pm\pi/a$, as is verified by FDTD simulations in Fig.~\ref{validation3}(a\&b). Since $H^{odd}_z(x)$ is an odd function, it always has a zero Fourier component at $k=0$. Therefore, dielectric-centered cavities (Fig.~\ref{schematics}) should have higher \emph{Q}-factors. However, in high-Q cavity designs, $\sigma a^2<<1$ is satisfied and thus both dielectric-centered and air-centered cavities have comparable \emph{Q}-factors. FDTD simulation shows that the above $H^{odd}_z$ and $H^{even}_z$ cavities have $Q_{\mathrm{tot}}=3.8\times10^8$ and $Q_{\mathrm{tot}}=3.5\times10^8$ respectively. The mode volume of the $H^{odd}_z$ cavity is $0.67(\lambda_{\mathrm{res}}/n_{\mathrm{Si}})^3$, smaller than the $H^{even}_z$ cavity ($V=0.76(\lambda_{\mathrm{res}}/n_{\mathrm{Si}})^3$).

The far field radiation patterns (obtained using FDTD simulations) of the two cavities are shown in Fig.~\ref{validation3}(c)\&(d). The powers, in both cases, are radiated at shallow angles ($>70^\circ$ zenith angle) to the direction of the waveguide. The $H^{odd}$ cavity has even less radiated power at small zenith angles, consistent with the above analysis. By integrating the zenith and azimuth angle dependent far field emission, we found that that 32\% and 63\% of the power emitted to $+\hat{z}$ direction can be collected by a NA=0.95 lens, respectively for $H^{odd}$ cavity and $H^{even}$ cavity.

\section{Ultra-high \emph{Q}, Dielectric-mode Photonic Crystal Nanobeam Cavities}
\subsection{Radiation-Q limited and waveguide-coupled cavities}
\begin{figure}
\centering
\includegraphics[width=12cm,height=4.5cm]{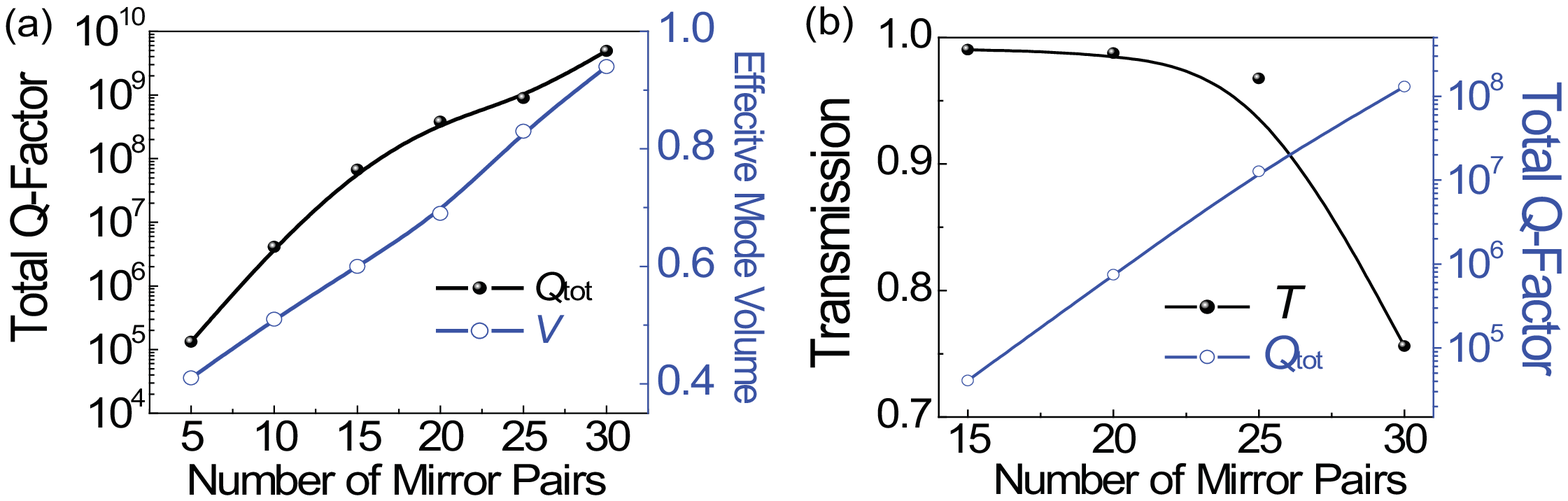}
\caption{\small
(a) Total \emph{Q}-factors (log(10) scale) and effective mode volumes ($V/(\lambda_{\mathrm{res}}/n_{\mathrm{Si}})^3$) of nanobeam cavities for different total number of mirror pair segments in the Gaussian mirror. In each case, 10 additional mirror segments with f=0.1 (maximum mirror strength)
are added on both ends of the Gaussian mirror. Therefore, the
total-\emph{Q} of the cavity is limited by radiation-\emph{Q}. A record ultra-high \emph{Q} of $5.0\times10^9$ is achieved with a Gaussian mirror that comprises 30 mirror segments and an
additional 10 mirror pairs on both ends.  (b) On-resonance transmissions and total \emph{Q}-factors (log(10) scale)  v.s the total number of mirror pair segments in the Gaussian mirror. In this case additional mirror pairs (10 of them) are not included. A record high-\emph{T} (97\%) and high-\emph{Q} ($1.3\times10^7$) cavity is achieved at $N=25$. }\label{3Ddie}
\end{figure}

Since the dielectric-centered $H^{odd}_z$ cavity has smaller \emph{V} than the $H^{even}_z$ one, we focus our discussion in the $H^{odd}_z$ case. Using the above design algorithm, we designed the Gaussian mirror and put 10 additional mirrors with the maximum mirror strength on both ends of the Gaussian mirror to obtain the radiation-limited cavity ($Q_{\mathrm{wg}}>>Q_{\mathrm{tot}}$). We find in Fig.~\ref{3Ddie} that $Q_{\mathrm{tot}}$ increases exponentially and \emph{V} increases linearly to the total number of mirror pairs in the Gaussian mirror ($N$). A record ultra-high \emph{Q} of $5.0\times10^9$ was achieved while maintaining the small mode volume of  $0.9\times (\lambda_{\mathrm{res}}/n_{\mathrm{Si}})^3$ at $N=30$.

\begin{figure}
\centering
\includegraphics[width=10.5cm,height=11.5cm]{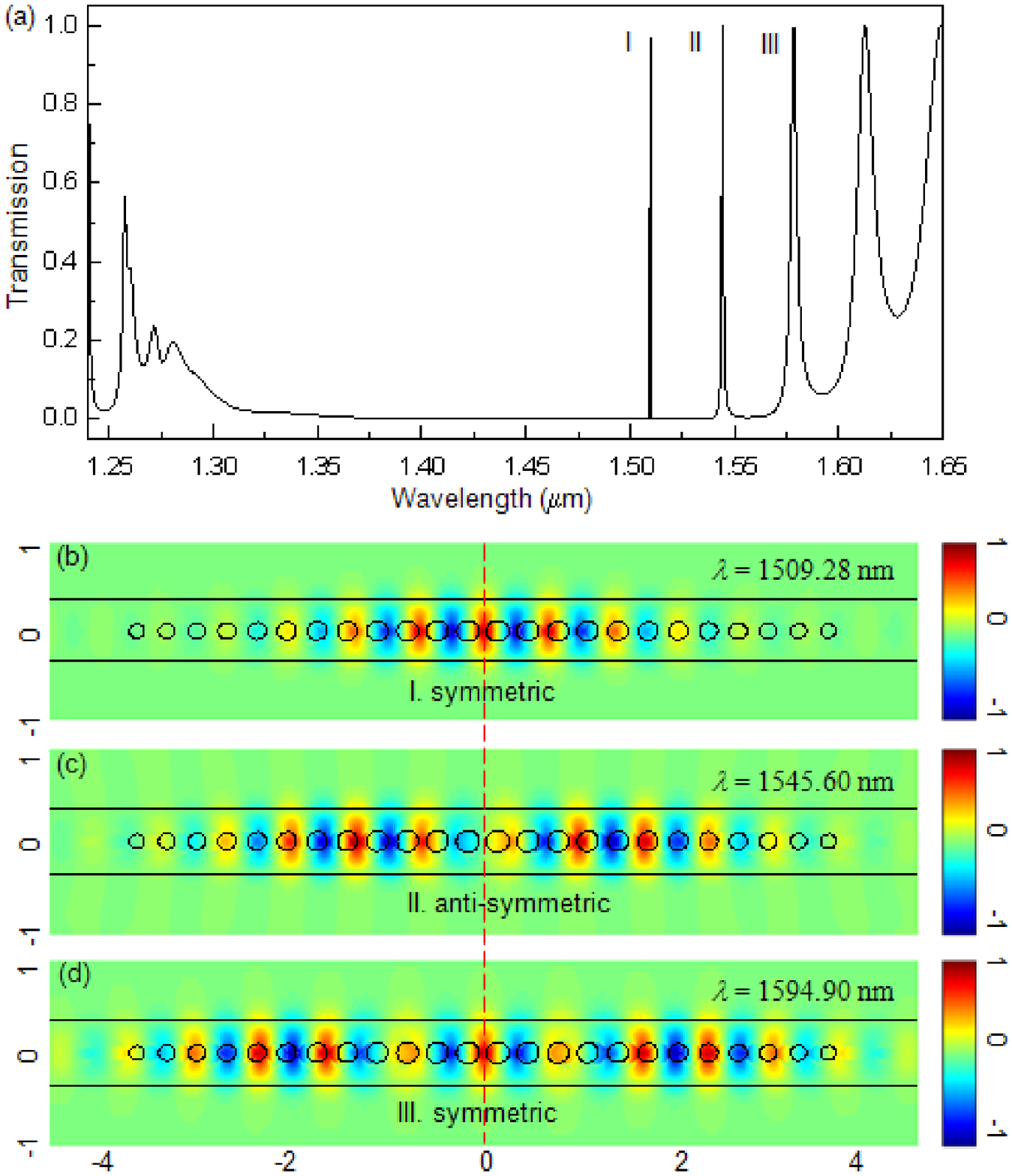}
\caption{\small (a) Transmission spectrum of the cavity from FDTD simulation. (b)-(d) The $E_y$ field distribution in the middle plain of the nanobeam cavity. Resonances and symmetries of the modes are indicated in the plot. Symmetry plane is perpendicular to the nanobeam direction, in the middle of the cavity, as indicated by the dashed line. Length unit in (b)-(d) is $\mu$m. } \label{3Ddiemodeanalysis}
\end{figure}

Our design
strategy has an additional important advantage over other types of
photonic crystal cavities\cite{Noda03}-\cite{Deotare08}, that is: the cavity naturally couples to the feeding waveguide, as the hole radii decrease away from the center of the cavity. High-\emph{Q} and high transmissions ($T$) cavities are possible with the above design steps (i)-(ix), with no additional "coupling sections" needed. We study $T$ and $Q_{\mathrm{total}}$ dependence on the total number of mirror pair segments in the Gaussian mirror ($N$) in Fig.~\ref{3Ddie}(b). Partial \emph{Q}-factors ($Q_{\mathrm{rad}},Q_{\mathrm{wg}}$) were obtained from FDTD simulations, and $T$ was obtained using $T=Q^2_{\mathrm{total}}/Q^2_{\mathrm{wg}}$\cite{joannopoulosbook}. As shown in Fig.~\ref{3Ddie}(b), we achieved a nanobeam cavity with $Q=1.3\times 10^7, T=97\%$ at $N=25$.

\subsection{Higher order modes of the dielectric-mode cavity}
The ultra-high \emph{Q} mode that we deterministically designed is the fundamental mode of the cavity. Meanwhile, higher order cavity modes also exist. The number of higher order modes depends on the width of the photonic band gap and total number of mirror segments in the Gaussian mirror. Fig.~\ref{3Ddiemodeanalysis}(a) shows the transmission spectrum of a waveguide-coupled nanobeam cavity, that has 12 mirror pair segments in the Gaussian mirror. Since the cavity was designed to be a waveguide-coupled cavity, the simulation was performed by exciting the input waveguide with a waveguide mode, and monitoring the  transmission through the cavity at the output waveguide. The band-edge modes are observed at wavelengths longer than 1.6$\mu$m and shorter than 1.3$\mu$m. Fig.~\ref{3Ddiemodeanalysis}(b)-(d) shows the major field-component ($E_y$) distribution of the three cavity modes. As expected, the eigenmodes alternate between symmetric and anti-symmetric modes. Symmetry plane is defined perpendicular to the beam direction, in the middle of the cavity (dashed line in Fig.~\ref{3Ddiemodeanalysis}). We note that transversely odd modes are well separated from the transversely symmetric cavity modes, hence were not considered in Fig.~\ref{3Ddiemodeanalysis}.

\section{Ultra-high \emph{Q}, Air-mode Photonic Crystal Nanobeam Cavities}
\subsection{Radiation-Q limited cavity}
\begin{figure}
\centering
\includegraphics[width=14cm,height=4.2cm]{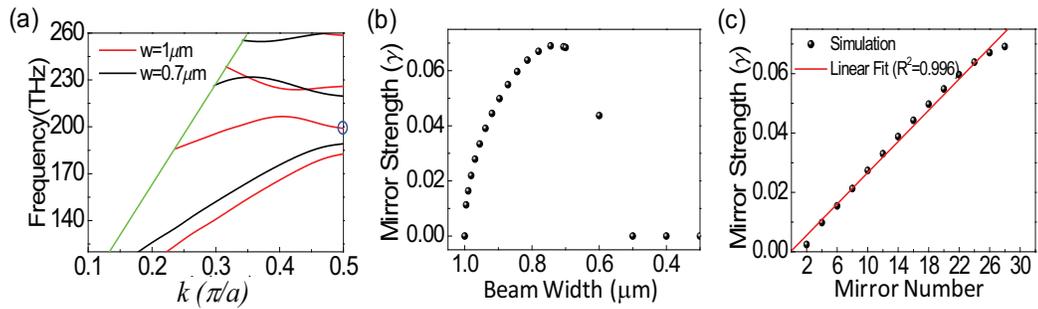}
\caption{\small
(a) TE band diagram for an air-mode nanobeam cavity. Hole radii $r=100\mathrm{nm},
a=330\mathrm{nm}$, b=1$\mu$m (red) and b=0.7$\mu$m (black). (b) Mirror strengths for different beam widths. (c) Linearization of mirror strengths after quadratic tapering the beam widths.}\label{3Dairmode}
\end{figure}
An air-mode cavity concentrates the optical energy in the low index region of the cavity. Therefore, these cavities are of interest for applications where strong interactions between light and material placed in the low index region of the cavity is required, including nonlinear enhancement\cite{leuthold10},  optical trapping\cite{greier03}, biochemical sensing\cite{psaltis06} and light-atom interaction\cite{vuckovic01}. The ultra-high \emph{Q} air-mode nanobeam cavity is realized by pulling the air-band mode of photonic crystal into its bandgap, which can also be designed using the same design principles that we developed for dielectric-mode cavities. In contrast to the dielectric-mode case, the resonant frequency of the air-mode cavity is determined by the air band-edge frequency of the center mirror segment. And to create the Gaussian confinement, the bandgaps of the mirror segments should shift to higher frequencies as they go away from the center of the cavity. This can be achieved by progressively increasing the filling fractions of the mirror segments away from the center of the structure (instead of decreasing in the dielectric-mode cavity case). One way to accomplish this is to increase the size of the holes away from the center of the cavity. While this may be suitable for non-waveguide coupled (radiation-\emph{Q} limited) cavities, it is not ideal for a waveguide-coupled cavity, where high transmission efficiency through the cavity is required. For this reason, we employ the design that relies on tapering of the waveguide width instead of the hole size. Similar geometry was recently proposed by Ahn et. al.\cite{ahn10} for the design of a dielectric-mode photonic crystal laser with different design method.

The same design steps can be followed as in the dielectric-mode cavity case, with the following changes: First, the adjusted frequency (198THz) is 1\% lower than the target frequency (200THz). (The thickness of the nanobeam is 220nm and period is 330nm, same as previous case.) Second, the nanobeam width at the center of the cavity is $w_{\mathrm{start}}=1\mu$m (Fig.~\ref{3Dairmode}(a)), with the hole radii kept constant at 100nm. Third, to create the Gaussian mirror, the beam widths are quadratically tapered from $w_{\mathrm{start}}=1\mu$m to $w_{\mathrm{end}}=0.7\mu$m, which produces the maximum mirror strength (band diagrams shown in Fig.~\ref{3Dairmode}(a)). This procedure involves calculating the mirror strength for several beam widths (Fig.~\ref{3Dairmode}(b)), each takes one or two minutes on a laptop computer. As shown in Fig.~\ref{3Dairmode}(c), the mirror strengths are linearized after the quadratic tapering. In order to achieve a radiation-\emph{Q} limited cavity, 10 additional mirror segments are placed at both ends of the Gaussian mirror that has beam width $w_{\mathrm{end}}=0.7\mu$m.

Similar in the dielectric-mode cavity cases, $H^{odd}_z$ and $H^{even}_z$ air-mode cavities can be formed by placing the air and dielectric in the central symmetric plane of the cavity, respectively. Again, we will focus on $H^{odd}_z$, air-mode cavities and the conclusions will be valid to the $H^{even}_z$ cavities as well. Fig.~\ref{3Dair}(a) shows the total \emph{Q} of nanobeam cavities hat have different total number of mirror pair segments in the Gaussian mirrors. We have achieved a record ultra-high \emph{Q} of $1.4\times10^9$, air-mode nanobeam cavity. As shown in Fig.~\ref{3Dair}(a), the effective mode volumes of the air-mode cavities are much larger than the dielectric-mode cavities.

\subsection{Cavity strongly coupled to the feeding waveguide}
\begin{figure}
\centering
\includegraphics[width=13cm,height=4.5cm]{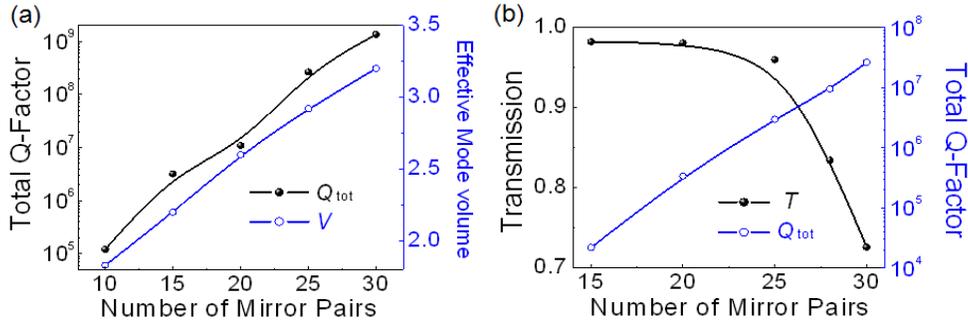}
\caption{\small
(a) Total \emph{Q}-factors (log(10) scale) and effective mode volumes ($V/(\lambda_{\mathrm{res}}/n_{\mathrm{Si}})^3$) of the nanobeam cavities for different total number of mirror pair segments in the Gaussian mirror.
In each case, 10 additional mirror segments with w=0.7$\mu$m are added on both ends of the Gaussian mirror,
so that the total-\emph{Q} of the cavity is limited by radiation-\emph{Q}. A record ultra-high \emph{Q} of $1.4\times10^9$ is achieved with a Gaussian mirror that comprises 30 mirror segments and 10 additional mirror pairs on both ends. (b) On-resonance transmissions and total \emph{Q}-factors (log(10) scale) v.s the total number of mirror pair segments in the Gaussian mirror. In this case additional mirror pairs (10 of them) are not included. A record high-\emph{T} (96\%) and high-\emph{Q} ($3.0\times10^6$) cavity is achieved at $N=25$.} \label{3Dair}
\end{figure}
As we have pointed out, the tapering-width approach (as compared to taping hole radii) offers a natural way of coupling the nanobeam air-mode cavity to the feeding waveguide. Since the width of the beam is decreasing, the cavity naturally couples to the feeding waveguide. We study $T$ and $Q_{\mathrm{total}}$ dependence on the total number of mirror pair segments in the Gaussian mirror ($N$) using FDTD simulations. As shown in Fig.~\ref{3Dair}(b), we were able to design a nanobeam cavity with $Q=3.0\times 10^6, T=96\%$ at $N=25$.

\subsection{Higher order modes of the air-mode cavity}
\begin{figure}
\centering
\includegraphics[width=10.5cm,height=8.5cm]{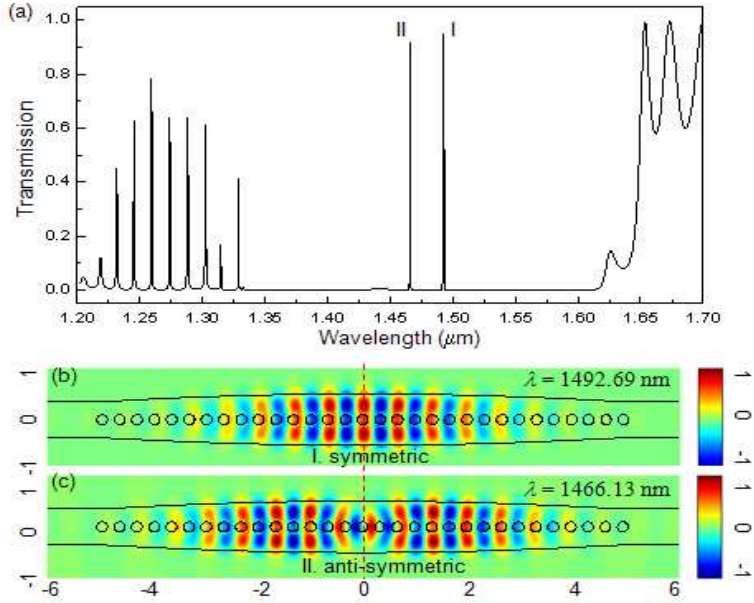}
\caption{\small (a) Transmission spectrum of the cavity from FDTD simulation. (b)\&(c) The $E_y$ field distribution in the middle plain of the nanobeam cavity. Resonances and symmetries of the modes are indicated in the plot. Symmetry plane is perpendicular to the nanobeam direction, in the middle of the cavity, as indicated by the dashed line. Length unit in (b)\&(c) is $\mu$m.} \label{3Dairmodeanalysis}
\end{figure}
The ultra-high \emph{Q} cavity that we were able to design is the fundamental mode of the cavity. Higher order modes coexist with the fundamental modes inside the band gap. Fig.~\ref{3Dairmodeanalysis}(a) shows the transmission spectrum of a waveguide-coupled air-mode nanobeam cavity, that has 15 mirror pair segments in the Gaussian mirror. The band-edge modes are observed at wavelengths longer than 1.6$\mu$m. The modes in the range of $1.2\mu$m to $1.35\mu$m are formed by the higher order band modes in Fig.~\ref{3Dairmode}(a). Fig.~\ref{3Dairmodeanalysis}(b)-(c) shows the major field-component distribution ($E_y$) of the two cavity modes inside the bandgap.

\section{Conclusion}
We have presented a detailed analysis and a deterministic design of the
ultra-high \emph{Q} photonic crystal nanobeam cavities. With this method, $Q>10^9$ radiation-limited cavity, and $Q>10^7, T>95\%$ waveguide-coupled cavity were deterministically designed. These \emph{Q}-factors are comparable with those found in whispering gallery mode (WGM) cavities\cite{Vernooy98}-\cite{adibi} and have ultra-small mode volumes, typically two or three orders of magnitude smaller than WGM ones. Furthermore, energy maximum can be localized in either the dielectric region or air region with this method. Although we demonstrated designs for TE-like, transversely symmetric cavity modes, the design method is universal, and can be applied to realize nanobeam cavities that support TM-polarized modes, as well as line-defect 2D photonic crystal cavities. We believe that the proposed method will greatly ease the processes of high \emph{Q} nanobeam cavity design, and thus enable both fundamental studies in strong light and matter interactions, and practical applications in novel light sources, functional optical components (filters, delay lines, sensors) and densely integrated photonic circuits.

\section{Acknowledgments}
We acknowledge numerous fruitful discussions with M.W. MuCutcheon and P. B.
Deotare. This work is supported by NSF Grant No. ECCS-0701417 and
NSF CAREER grant.

\end{document}